\documentclass[a4paper,11pt]{article}
\usepackage[T1]{fontenc}
\usepackage[utf8x]{inputenc}
\usepackage{color}
\usepackage{bm}
\usepackage{stmaryrd, mathrsfs, dsfont, amsmath, latexsym}
\usepackage{youngtab}
\usepackage{hyperref}
\usepackage{verbatim}
\usepackage{cite}
\usepackage{authblk}
\numberwithin{equation}{section}

\def \de{\partial}

\def \H{{\cal H}}
\def \J{{\cal J}}
\def \Y{{\cal Y}}
\def \N{{\cal N}}

\def \D{{\cal D}}
\def \Tr{{\rm Tr}}
\def \gTr{{{\cal T}\!r}} 
\newcommand{\ket}[1]{\left\vert #1 \right\rangle}
\newcommand{\bra}[1]{\left\langle #1 \right\vert}
\newcommand{\braket}[2]{\left\langle #1\,,#2\right\rangle}

\begin{document}

\begin{titlepage}

\setcounter{page}{1}
\begin{flushright}
LMU-ASC 42/18\\
\end{flushright}
\begin{center}

\vskip 1cm


{\Large \bf  Einstein gravity from the ${\N=4}$ spinning particle}


\vspace{12pt}

R. Bonezzi\,$^{1,2}$, A. Meyer\,$^3$, I. Sachs\,$^3$\\

\vskip 25pt

{\em $^1$ \hskip -.1truecm Physique th\'eorique et math\'ematique, 
Universit\'e de Mons -- UMONS \\
20, Place du Parc, 7000 Mons, Belgium\vskip 5pt }

\vskip 10pt
{\em $^2$ \hskip -.1truecm Dipartimento di Fisica, Universit\`a di Bologna and INFN sezione di Bologna,\\ via Irnerio 46,
I-40126 Bologna, Italy}

\vskip 10pt
 {\em $^3$ \hskip -.1truecm Arnold Sommerfeld Center for Theoretical Physics,
Ludwig-Maximilian-Universit\"at ,\\
Theresienstr. 37, D-80333 M\"unchen, Germany}

\end{center}

\vskip 15pt

\begin{center}

{emails: \small{{\tt roberto.bonezzi@umons.ac.be},  
{\tt Adiel.Meyer@physik.uni-muenchen.de}, 
{\tt Ivo.Sachs@physik.uni-muenchen.de}}}
\end{center}

\vskip .5cm

\begin{abstract}
    We obtain a manifestly background independent BRST quantization of the $\N=4$ supersymmetric spinning particle. We show that nilpotency of the BRST charge $Q$ implies the Einstein equations admitting a cosmological constant of indefinite sign. The physical graviton states are given by the vertex operator, obtained by an infinitesimal variation of $Q$, acting on diffeomorphism ghost states. In addition, the tree-point graviton scattering vertex is correctly reproduced by the worldline computation with these vertex operators. 
\end{abstract}

\end{titlepage}

\newpage

\tableofcontents

\vspace{1 cm}

\section{Introduction}
There are several reasons for considering the worldline approach to field theory and gravity. For one thing it provides a first-quantized description of spin one fields \cite{Barducci:1976xq,Brink:1976uf}, that allows for a simple mechanism to compute one-loop one particle irreducible actions (1PI) for Yang-Mills theory  \cite{Strassler:1992zr,Reuter:1996zm,Sato:1998sf,Schubert:2001he,Bastianelli:2013pta}. Similarly, trace anomalies and gravitational one-loop effective actions were derived in the worldline formalism for scalar and spinor loops in \cite{Bastianelli:1991be,Bastianelli:1992ct,Bastianelli:2002fv,Bastianelli:2002qw} and for loops of differential forms in \cite{Bastianelli:2005vk,Bastianelli:2005uy}. 
Furthermore, the field theory ghosts  are automatically taken care of by the worldline ghosts, since gauging the worldline supersymmetries amounts, as in string theory, to removing unphysical degrees of freedom in spacetime. Indeed,
at tree-level the BRST quantization of the spinning particle naturally produces a BV-action in space time \cite{Thorn:1988hm,Barnich:2003wj,Barnich:2004cr,Barnich:2006pc}. 
The fields and anti-fields of variable degree are included by simply relaxing the worldline ghost number. 
String field theory is obtained in the same way upon replacing the worldline by a world sheet \cite{Witten:1985cc,Zwiebach:1992ie}. Even for the massless particle, introducing a world sheet as the complexification of the worldline has advantages since it allows to reorganize the Feynman amplitudes in an efficient way much simplifying the calculation of scattering amplitudes using world sheet methods \cite{Bern:1991aq} and more recently for tree-level amplitudes in ambitwistor strings \cite{Mason:2013sva,Adamo:2014wea}.    

On the other hand, the construction of a manifestly background invariant action along this line has been a major obstacle in string field theory. There are various reasons for this: One problem is that we do not know how to couple massive string states to the world sheet of the string at the non-linear level. In fact, even for the massless fields such as the graviton and the dilaton the absence of conformal invariance in a general background renders the construction of the BRST charge problematic (see e.g. \cite{Mansfield:1987ka} for an attempt in this direction). These problems should be absent for the worldline where neither massive states nor conformal invariance pose a problem. However, in the case of self-interacting theories, even for the spinning particle with non-zero spin the BRST charge fails to square to zero on shell unless suitable constraints are imposed on the representation space. Unlike for the string, the truncation of the Fock space becomes possible for the spinning particle, due to the  $R$-symmetry that comes with the extended worldline supersymmetry. In \cite{Dai:2008bh} this program was carried out successfully for Yang-Mills theory described by a worldline with $\N=2$ SUSY\footnote{In general, spinning particles with $\N$ supersymmetries describe spin $\frac{\N}{2}$ particles in spacetime \cite{Berezin:1976eg,Gershun:1979fb,Henneaux:1987cp,Howe:1988ft,Siegel:1988ru,Bastianelli:2007pv,Bastianelli:2008nm,Corradini:2010ia,Bastianelli:2012bn}}. There, the BRST charge was constructed for an arbitrary Yang-Mills background and the field equations for the latter were recovered from the nilpotency of $Q$ on a suitably restricted Fock space with fixed $U(1)$ $R$-charge, still big enough to contain all physical degrees of freedom. Furthermore, it was shown that the variation of $Q$ on a solution of the field equations reproduces a vertex operator for the gluon that produces the physical state when acting on the Yang-Mills ghost vacuum. 

The theory just described can be coupled to off-shell gravity with no further conditions \cite{Bastianelli:2005vk}. However, the consistent coupling of the worldline with ${\N=4}$ extended SUSY, needed to include the graviton as a propagating degree of freedom, was so far lacking\footnote{In \cite{Bastianelli:2013tsa} a worldline approach was proposed to describe Einstein gravity at one-loop, but the gauge structure of the graviton was treated somehow ad hoc directly from the field theory.}. This is the problem we propose to solve in this note. In particular, we will identify the correct constraint on the Fock space consistent with the field equations in space-time\footnote{The fact that coupling the ${\N=2}$ 
worldline to gravity is automatically consistent while ${\N=4}$ is not, is easy to understand from the spacetime perspective. Indeed, gluons can propagate in any geometry while gravitons can propagate only on Einstein spacetimes.}. As usual, at quantum level we should impose only half of the constraints on the Fock space. For the $SO(4)$ $R$-symmetry this can be done by choosing a decomposition of the $so(4)$ Lie-algebra that maintains manifest covariance only under a $u(2)$-subalgebra. With this restriction the graviton is the only propagating degree of freedom in this theory. In this context it is worthwhile to point out that the truncation of the massless NS-spectrum of string theory to the pure graviton sector is possible for the worldline, while it is not the case for string theory, due to the the enhancement of the $R$-symmetry on the worldline.  

Nilpotency of $Q$ then requires the background to be Einstein allowing, in particular, for a cosmological constant of indefinite sign, in agreement with expectations from General Relativity. Consistent, that is, nilpotent infinitesimal deformations $Q=Q_0+V$ of a classical background are given by equivalence classes in the adjoint cohomology of $Q_0$ and should thus be isomorphic to the physical states on this background. We find indeed that upon acting with such $V$ on suitable diffeomorphism ghost states generates the physical graviton states in the Fock space in analogy with \cite{Dai:2008bh} for ${\N=2}$. 
  
The organization of this paper is as follows: In section 2 we describe the worldline theory of the ${\N=4}$ spinning particle, together with the possible constraints that can be imposed on the representation space. In section 3 we review the Dirac  quantization and describe the physical spectrum in flat space, both, in terms of  the linearized curvatures as well as the potentials. The two descriptions are related by a shift in the normal ordering constant of a $U(1)$ $R$-current. In section 4 we focus on the BRST quantization. In section 5 we construct the BRST charge in a curved metric background and analyze the field equations implied by the nilpotency of $Q$. In section 6 we obtain a vertex operator for the graviton by variation of the BRST charge w.r.t. the background metric and determine the appropriate ghost state to evaluate the corresponding scattering amplitudes. Some technical details are referred to the appendices. In particular, the $R$-symmetry enhancement to $SO(4)$ is explained in appendix \ref{string} in term of a twisted dimensional reduction of the world sheet action. 

\section{${\N=4}$ supersymmetric spinning particle}

Let us start by reviewing the (quantized) graded phase space of the point particle. The canonical coordinates are $(x^\mu, p_\mu, \Theta^\mu_I)\,$, with $\mu=1,..,d$ a flat spacetime index and $I=1,..,4$ an internal index. They are subject to the commutation relations
\begin{equation}
[x^\mu, p_\nu]=i\,\delta^\mu_\nu\;,\quad \{\Theta^\mu_I, \Theta^\nu_J\}=\delta_{IJ}\,\eta^{\mu\nu}   \;. 
\end{equation}
The four hermitian supercharges $Q_I:=\Theta_I\cdot p$ together with the hamiltonian $H:=\tfrac12\, p^2\equiv-\tfrac12\Box$ generate the ${\N=4}$ worldline supersymmetry algebra
\begin{equation}
\{Q_I, Q_J\}=2\,\delta_{IJ}\,H\;,\quad [Q_I, H]=0\;,
\end{equation}
with manifest $so(4)$ $R$-symmetry algebra generated by $J_{IJ}:=i\,\Theta_{[I}\cdot\Theta_{J]}$ obeying
\begin{equation}
[J_{IJ}, Q_K]=2i\,Q_{[I}\,\delta_{J]K}\;,\quad [J_{IJ}, J_{KL}]=4i\,\delta_{[K[J}\,J_{I]L]}\;.    
\end{equation}
The Hilbert space for the fermionic algebra is  generated from the oscillator variables (omitting spacetime indices) $\theta_i:=\tfrac{1}{\sqrt2}(\Theta_i+i\Theta_{i+2})$ and $\bar\theta^i:=\tfrac{1}{\sqrt2}(\Theta_i-i\Theta_{i+2})$, $i=1,2$, obeying
\begin{equation}
\{\bar\theta^i_\mu, \theta_j^\nu\}=\delta^i_j\,\delta^\nu_\mu\;,\quad \{\bar\theta^i_\mu, \bar\theta^j_\nu\}=0=\{\theta_i^\mu, \theta_j^\nu\} \;.   
\end{equation}
By choosing a Fock vacuum annihilated by $\bar\theta_\mu^i\,$, an arbitrary state $\ket{\psi}$ in the full Hilbert space is isomorphic to the wave function 
\begin{equation}\label{wavefunction}
\psi(x,\theta_i)=\sum_{m,n=0}^d\psi_{\mu[m]\vert \nu[n]}(x)\,\theta_1^{\mu_1}...\theta_1^{\mu_m}\,\theta_2^{\nu_1}...\theta_2^{\nu_n}\sim\bigoplus_{m,n}\;m\left\{\,\yng(1,1,1,1)\right.\,\otimes\, n\left\{\yng(1,1,1)\,\right.
\end{equation}
where the fermions $\bar\theta^i_\mu$ act as $\frac{\de}{\de\theta_i^\mu}\,$. Above we have displayed the spacetime tensor field content by the Young diagrams on the right hand side. We used the condensed notation for antisymmetrized indices $\mu[m]:=[\mu_1...\mu_m]$ and a vertical bar to separate  indices with no symmetry relations.
The supercharges act on wave functions as antisymmetrized gradients and divergences: 
\begin{equation}
q_i=-i\,\theta_i^\mu\de_\mu\;,\quad \bar q^i=-i\,\de^\mu\frac{\de}{\de\theta_i^\mu}\;,    
\end{equation}
where the redefinition from $Q_I$ to $(q_i,\bar q^i)$ follows immediately from the definitions of $\theta_i^\mu$ and $\bar\theta^i_\mu\,$. The adjoint operation is defined by the inner product
\begin{equation}\label{WaveFunc}
\braket{\phi}{\psi}:=\int d^dx\,\big[\phi^*(x,\de_{\theta_i})\,\psi(x,\theta_i) \big]\rvert_{\theta_i=0} \;,  
\end{equation}
which gives $\bar q^i=(q_i)^\dagger\,$. The $so(4)$ generators split under the $\Theta_I\rightarrow(\theta_i, \bar\theta^i)$ redefinition maintaining only manifest covariance under a $u(2)$ subalgebra: $J_{IJ}\rightarrow(J^i_j, J_{ij}, \bar J^{ij})$ with explicit realization 
\begin{equation}\label{so4pack}
J^i_j=
\begin{pmatrix}
N_1-\tfrac{d}{2} & Y^\dagger\\ Y & N_2-\tfrac{d}{2}
\end{pmatrix}\;,\quad
J_{ij}=
\begin{pmatrix}
0 & {\rm g}\\ -{\rm g} & 0
\end{pmatrix}\;,\quad 
\bar J^{ij}=
\begin{pmatrix}
0 & {\rm Tr}\\ -{\rm Tr} & 0
\end{pmatrix}
\end{equation}
where
\begin{equation}
\begin{split}
& N_i:=\theta_i\cdot\frac{\de}{\de\theta_i}\;,\quad\text{$i$ not summed, counts indices in column $i$}  \\
& Y:= \theta_1\cdot\frac{\de}{\de\theta_2}\;,\quad\text{Young antisymmetrizer $2\to1$}\;,\quad Y^\dagger:= \theta_2\cdot\frac{\de}{\de\theta_1}\;,\quad\text{Young antisymmetrizer $1\to2$}\\
& {\rm g}:=\theta_1\cdot\theta_2\;,\quad\text{insertion of the metric $\eta_{\mu\nu}$}\;,\quad {\rm Tr}:=\frac{\de^2}{\de\theta_1\cdot\de\theta_2}\;,\quad \text{trace between columns.} 
\end{split}    
\end{equation}
In order to describe relativistic (massless in the case at hand) particles, the hamiltonian has to be gauged in order to ensure the mass-shell condition $p^2\approx0\,$. From the worldline viewpoint it is also clear that the supersymmetries should be gauged in order to have unitarity. Indeed, in light-cone gauge the local worldline supersymmetries precisely get rid of the light-cone polarizations $\Theta^\pm_I$ and allow to construct a manifest unitary spectrum out of transverse oscillators.
The situation for the $R$-symmetries leaves a much wider choice. As it can be seen from the operators above, the $R$-symmetry generators perform algebraic operations on the spacetime tensors. The larger $so(4)$ subalgebra is gauged, the less reducible the spacetime spectrum is. It is worth to notice that the shift $-\tfrac{d}{2}$ in the definition of $J_i^i$ is a quantum ordering effect. The value $-\tfrac{d}{2}$ is the only one that does not introduce a central extension in the $so(4)$ algebra. This condition can be relaxed if we impose that only a suitable subalgebra annihilates physical states. In this note we will consider only the case of maximal gauging of the $R$-symmetry, namely the full $so(4)\,$, that yields an irreducible spectrum containing only the graviton as physical state.
The classical worldline action describing the model, the ${\cal N}=4$ spinning particle, reads
\begin{equation}
S=\int d\tau \big[p_\mu\dot x^\mu+i\bar\theta^i_\mu\dot\theta^\mu_i-\tfrac{e}{2}\,p^2-i\chi_i\,\bar\theta^i\!\cdot p-i\bar\chi^i\,\theta_i\!\cdot p-a^{IJ}\,J_{IJ}\big]\;,    
\end{equation}
where $e(\tau)$ is the worldline einbein gauging $p^2\,$, and playing the role of gauge field for 1D reparametrization invariance. Correspondingly, the four supersymmetries are gauged by worldline gravitini $\chi_i(\tau)$ and $\bar\chi^i(\tau)\,$, while $so(4)\,$, generated by $J_{IJ}\,$, is gauged by the one-dimensional Yang-Mills field $a^{IJ}(\tau)\,$.

\section{Dirac quantization}

In this section we review the Dirac quantization of the model by first assuming that the quantum ordering of the operators $J_i^i$ does not introduce central terms in the $so(4)$ algebra, \emph{i.e.} $J_i^i=N_i-\tfrac{d}{2}\,$. This yields the physical spectrum in terms of linearized curvatures obeying first order differential equations. By changing the constant shift in the quantum operators $J^i_i$ it is possible to describe the degrees of freedom in terms of gauge fields, that are necessary in order to introduce self-interactions. This second option will be described in section \ref{gaugedirac}.

\subsection{Curvature description}

We now proceed to review the Dirac quantization when the entire $R$-symmetry algebra $so(4)\,$ is gauged \cite{Bastianelli:2008nm}.
In this case all the constraints can be imposed at the quantum level on the physical state $\ket{R}\,$. An independent set is given by\footnote{The constraint $\bar q^i$ is automatically satisfied thanks to the $(q_i, {\rm Tr})$ algebra, $H$ consequently follows, and ${\rm g}\ket{R}=0$ is equivalent to the trace constraint upon double dualization.}
\begin{equation}
\left(N_i-\tfrac{d}{2}\right)\ket{R}=0=Y\ket{R}\;,\quad q_i\ket{R}=0\;,\quad {\rm Tr}\ket{R}=0    
\end{equation}
We stress that the shifts on the number operators $N_i$ are the only ones that preserve the full $so(4)$ at the quantum level. Hence, this model describes a graviton only in $d=4\,$, to which we shall restrict at the moment. 
The first set of constraints imposes $gl(d)$ irreducibility of the spacetime tensor: after imposing $(N_i-\tfrac{d}{2})\ket{R}=0$ in four dimensions one has
\begin{equation}\label{Curvature diagrams}
\ket{R}\;\sim\; \yng(1,1)\otimes\yng(1,1) = \yng(2,2)\;\oplus\;\yng(2,1,1)\;\oplus\;\yng(1,1,1,1)   \quad,
\end{equation}
where the first diagram corresponds to the spin two Riemann curvature.
By enforcing the Young constraint $Y\ket{R}=0$ the last two components of \eqref{Curvature diagrams} are projected out, and one is left with a field with the algebraic symmetries of the Riemann tensor:
\begin{equation}
R(x,\theta_i)=R_{\mu\nu\lambda\sigma}(x)\,\theta^\mu_1\theta^\nu_1\theta^\lambda_2\theta^\sigma_2 \; \sim\;\yng(2,2)  
\end{equation}
subject to the other constraints that are integrability and tracelessness conditions
\begin{equation}\label{spin2curvatureeoms}
\de_{[\mu}R_{\nu\lambda]\sigma\rho}=0\;,\quad R^\lambda{}_{\mu\lambda\nu}=0\;,    
\end{equation}
that play the role of equations of motion\footnote{The same geometric field equations in the context of higher-spin gauge theories were derived in \cite{Bekaert:2002dt,Bekaert:2003az,Bekaert:2006ix}.}.
At this stage one can analyze these equations purely in terms of curvature. On-shell one has $\Box R_{\mu\nu\lambda\sigma}=0\,$, that allows to choose a light-cone frame where only $p_+$ is nonzero. Solving the field equations \eqref{spin2curvatureeoms} the only non-vanishing components of the curvature are $R_{+i+j}\,$, traceless in the transverse indices $ij\,$, that propagate massless spin two degrees of freedom in terms of the linearized Weyl tensor. Alternatively, one can solve the integrability condition by integrating in the gauge potential as \begin{equation}
R_{\mu\nu\lambda\sigma}=4\,\de_{[\mu}\de_{[\lambda}h_{\sigma]\nu]}\quad\longleftrightarrow\quad \ket{R}=q_1q_2\ket{h}\;,    
\end{equation} 
in which case the traceless curvature condition becomes Fierz-Pauli equation for the massless graviton $h_{\mu\nu}\,$, \emph{i.e.}
\begin{equation}
\Box h_{\mu\nu}-2\,\de_{(\mu}\de\cdot h_{\nu)}+\de_\mu\de_\nu h^\lambda{}_\lambda =0   \;, 
\end{equation}
 that is nothing but the linearized Ricci tensor around flat space. The field equations in this form are clearly invariant under linearized diffeomorphisms (spin two gauge symmetry): $\delta h_{\mu\nu}=\de_{(\mu}\varepsilon_{\nu)}\,$, but we mention that spacetime gauge symmetry arises in this description only upon integrating in the potential $h_{\mu\nu}$ to solve the integrability condition, whereas the original equations in terms of curvatures have no gauge symmetry.

\subsection{Gauge field description}\label{gaugedirac}

The Dirac quantization reviewed in the last section seems to naturally describe the field content in a first-order gauge invariant formulation based on linearized curvatures. On the other hand, light-cone quantization and the covariant path integral display gauge fields in the spectrum, rather than curvatures. The two descriptions are clearly equivalent at the free level, while introducing interactions generally prevents the use of curvatures. 

To describe the above model in terms of potentials à la Dirac, one has to change the ordering shift\footnote{The two different normal-ordering constants, $\tfrac{d}{2}$ or $\tfrac{d-2}{2}\,$, depend on whether one prescribes to count the normal ordering of all fermions or, rather, only the transverse ones.} in the definition of $J^i_i$ to $J^i_i=N_i-\tfrac{d-2}{2}$ in any dimension. In this case only half of the supercharges can annihilate physical states, while the second half will generate null states, as it is customary in Gupta-Bleuler and string theory old covariant quantization. 
Similarly, only half (for conjugated pairs) of the $so(4)$ generators can annihilate physical states.
An independent set of constraints is then given by\footnote{We freely switch between $\Box$ and $H$ to denote the hamiltonian constraint.}
\begin{equation}
\left(N_i-\tfrac{d-2}{2}\right)\ket{h}=0\;,\quad \left(Y,{\rm Tr}\right)\ket{h}=0\;,\quad \bar q^i\ket{h}=0\;,\quad\Box\ket{h}=0  
\end{equation}
Although the above constraints form a closed subalgebra for any shift $J_i^i=N_i-n\,$, so that one may try to set $n=1$ in any dimension, the corresponding classical algebra is broken for $d\neq4$ and it is not clear how to perform the path integral. More precisely, the classical counterpart of the number operator constraint is $\theta_i\cdot\bar\theta^i=0$ for fixed $i\,$. The Dirac quantization admits then the two descriptions in terms of curvatures or gauge fields as
\begin{equation*}
\theta_i\cdot\bar\theta^i+\lambda=0\quad\overset{{\rm quantize}}{\longrightarrow}\left\{
\begin{array}{c}
(N_i-\tfrac{d}{2}+\lambda)\ket{R}=0\quad \text{gauge invariant curvature description}\\[3mm]
(N_i-\tfrac{d-2}{2}+\lambda)\ket{h}=0\quad \text{Gupta-Bleuler gauge field description}
\end{array}
\right.    
\end{equation*}
so that the desired field content corresponds to $\lambda=\tfrac{d}{2}-2\,$. If the entire $so(4)$ classical algebra is gauged, it suffers a classical anomaly for $\lambda\neq0\,$. One could try to avoid the problem by gauging the classical counterpart of ${\rm Tr}\,$, \emph{i.e.} $\bar\theta^2\cdot\bar\theta^1$ without gauging its conjugate\footnote{The classical anomaly arises from the Poisson bracket $\{\bar\theta^2\cdot\bar\theta^1, \theta_1\cdot\theta_2\}_{\rm PB}\,$.} $\theta_1\cdot\theta_2\,$, but this would break the reality of the classical action.

Restricting to four dimensions one recovers $J^i_i=N_i-1$ that, together with the Young constraint, is solved by the symmetric tensor $h(x,\theta_i)=h_{\mu\nu}(x)\,\theta_1^\mu\theta_2^\nu\,$. The remaining constraints $(\Tr,\bar q^i,\Box)\ket{h}=0$ then yield the Fierz-Pauli system
\begin{equation}
\Box h_{\mu\nu}=0\;,\quad \de^\mu h_{\mu\nu}=0\;,\quad h^\lambda{}_{\lambda}=0   \end{equation}
for massless spin two in partially gauge fixed form.
In the Dirac approach the presence of gauge symmetry manifests with the appearance of null states in the Hilbert space. These are physical states with zero norm and vanishing scalar product with all other physical states, that one can mod out from the physical spectrum. In the present case one can indeed see that fields of the form $h_{\mu\nu}=\de_{(\mu}\varepsilon_{\nu)}$ are null for $\varepsilon_\mu$ transverse and harmonic. Modding these out one is left with the transverse and traceless polarizations of the graviton $h_{ij}\,$.

\section{BRST quantization}\label{BRSTsection}

We shall now focus on the BRST quantization of the model, as it will be the starting point for introducing a curved background in the next section.
In the following, we will treat the $R$-symmetry $so(4)$ constraints and the SUSY constraints $(q_i,\bar q^i, \Box)$ on different footings. Namely, we will associate ghosts and BRST operator only to the superalgebra
\begin{equation}\label{susyalone}
\{q_i,\bar q^j\}=-\delta_i^j\,\Box\;,\quad \{q_i, q_j\}=0\;,\quad\{\bar q^i, \bar q^j\}=0\;,    
\end{equation}
corresponding to spacetime differential constraints. The algebraic $so(4)$ operators instead, (extended by appropriate ghost contributions,) will  be imposed separately as constraints on the BRST Hilbert space.
The reason to proceed this way is twofold: the introduction of $so(4)$ ghosts would result in a plethora of unnecessary auxiliary fields, and we will show that the present treatment is equivalent to Dirac quantization. The second, more important, reason is that the BRST quantization in curved space is consistent only as a cohomology on the constrained Hilbert space, as it will be shown in the next section.

We thus proceed by assigning the ghost-antighost canonical pair $(c,b)$ to the hamiltonian as well as bosonic superghost pairs $(\bar\gamma^i, \beta_i)$ and $(\gamma_i,\bar\beta^i)$ to the supercharges $q_i$ and $\bar q^i\,$, obeying canonical commutation relations
\begin{equation}
\{b,c\}=1\;,\quad [\beta_i,\bar\gamma^j]=\delta_i^j\;,\quad[\bar\beta^j,\gamma_i]=\delta_i^j\;,  
\end{equation}
with ghost number assignments ${{\rm gh}(c,\gamma_i,\bar\gamma^i)=+1}$ and ${{\rm gh}(b,\beta_i,\bar\beta^i)=-1}\,$.
The BRST differential associated to the algebra \eqref{susyalone} takes the form
\begin{equation}
Q:= c\,\Box+\gamma_i\,\bar q^i+\bar\gamma^i\, q_i+\bar\gamma^i\gamma_i\, b \;,\quad Q^2=0  \;. 
\end{equation}
With the hermiticity assignments $(\gamma_i)^\dagger=\bar\gamma^i$ and $(\beta_i)^\dagger=-\bar\beta^i\,$, the $b$ and $c$ ghosts being self-adjoint, one has $Q^\dagger=Q\,$.
In order to impose the $so(4)$ constraints on the BRST wave function we have to extend them by ghost contributions: $J_{IJ}\to \J_{IJ}$ as to commute with the BRST charge $Q\,$.
Explicitly, we have
\begin{equation}
\begin{split}
\J_i^j &:= \theta_i\cdot\bar\theta^j+\gamma_i\bar\beta^j-\bar\gamma^j\beta_i-\tfrac{d}{2}\,\delta_i^j \;,\\[2mm]
\gTr &:= \bar\theta^1\cdot\bar\theta^2+\bar\gamma^1\bar\beta^2 - \bar\gamma^2\bar\beta^1\;,\\[2mm]
{\cal G} &:= \theta_1\cdot\theta_2+\gamma_1\beta_2 - \gamma_2\beta_1\;,
\end{split}    
\end{equation}
where the $u(2)$ generators $\J_i^j$ correspond to the number operators for $i=j$ and to the Young antisymmetrizers for $i\neq j$ as in \eqref{so4pack}.

We choose the ghost vacuum $\ket{0}$ to be annihilated by $(b,\bar\gamma^i,\bar\beta^i)\,$, so that a general state $\ket{\Psi}$ in the BRST extended Hilbert space is isomorphic to the wave function
$\Psi(x,\theta_i\,|c,\gamma_i,\beta_i)\,$,
on which $(b,\bar\gamma^i,\bar\beta^i)$ are realized as $(\frac{\de}{\de c},-\frac{\de}{\de \beta_i}, \frac{\de}{\de \gamma_i})\,$. With the given choice of vacuum, the ghost number of the wave function is unbounded both from above and below, and the operator $Q$ takes the form 
\begin{equation}
Q=c\,\Box+\gamma_i\,\bar q^i-q_i\,\frac{\de}{\de\beta_i}-\gamma_i\,\frac{\de^2}{\de\beta_i\de c}\;. 
\end{equation}
As in any BRST system with a self-adjoint $bc$-ghost pair, one has
$\bra{0}\left.\!0\right\rangle=\bra{0}bc+cb\ket{0}=0$ and the non-vanishing inner product requires one insertion of the $c$ ghost:
\begin{equation}
\bra{0}c\ket{0}\sim1    \;.
\end{equation}
This fixes the ghost number of $\ket{0}$ to be $-\tfrac12\,$, but we usually remove this offset and count ghost number by assigning zero to $\ket0\,$.

The relevant $so(4)$ generators to be imposed as constraints on the BRST Hilbert space are the number operators $\J_i^i$ ($i$ not summed), the Young antisymmetrizer $\Y:=\J_1^2$ and the trace $\gTr$ that we will collectively denote ${\cal T}_\alpha:=(\J_i^i,\Y,\gTr)\,$. When acting on the wave function $\Psi$, they take the form
\begin{equation}
\begin{split}
\J_i^i &= N_{\theta_i}+N_{\gamma_i}+N_{\beta_i}-\tfrac{d-2}{2}=:\N_i-\tfrac{d-2}{2}\;, \\[2mm]
\Y &= \theta_1\cdot\frac{\de}{\de\theta_2}+\gamma_1\frac{\de}{\de\gamma_2}+\beta_1\frac{\de}{\de\beta_2}\;,\\[2mm]
\gTr &= \frac{\de^2}{\de\theta_1\cdot\de\theta_2}+\frac{\de^2}{\de\gamma_1\de\beta_2}-\frac{\de^2}{\de\gamma_2\de\beta_1}\;.
\end{split}    
\end{equation}
Let us stress that, thanks to the choice of vacuum, one has $\J_i^i=\N_i-1$ in four dimensions, and just imposing $(\N_i-1)\ket{\Psi}=0$ reduces the infinitely many\footnote{Recall that the ghosts $\gamma_i$ and $\beta_i$ are bosonic. We choose a polynomial basis for them, where order by order the operators $\N_i$ are well defined.} components of $\Psi$ to a handful of fields with a precise spacetime interpretation, as we shall see next.

Our BRST system is then defined by
\begin{equation}\label{BRSTsystem}
\begin{split}
& Q\Psi=0\;,\quad \delta\Psi=Q\Lambda\;,\\[2mm]
&{\cal T}_\alpha\Psi=0\;,\quad {\cal T}_\alpha\Lambda=0\;,
\end{split}    
\end{equation}
whose consistency is guaranteed by $[Q,{\cal T}_\alpha]=0\,$. This is equivalent to saying that we are studying the cohomology of $Q$ on the restricted Hilbert space defined by $\H_{\rm red}:=\ker{\cal T}_\alpha\,$. 
The most general state in this subspace can be written as
\begin{equation}\label{BVstringfield}
\begin{split}
\ker{\cal T}_\alpha\ni  \Psi(x,\theta_i\,|c,\gamma_i,\beta_i) &= h_{\mu\nu}(x)\,\theta^\mu_1\theta^\nu_2+\tfrac12\,h(x)\,(\gamma_1\beta_2-\gamma_2\beta_1)-\tfrac{i}{2}\,v_\mu(x)\,(\theta^\mu_1\beta_2-\theta^\mu_2\beta_1)c\\[1mm]
&-\tfrac{i}{2}\,\xi_\mu(x)\,(\theta^\mu_1\beta_2-\theta^\mu_2\beta_1)\\[2mm]
&+h^*_{\mu\nu}(x)\,\theta^\mu_1\theta^\nu_2c+\tfrac12\,h^*(x)\,(\gamma_1\beta_2-\gamma_2\beta_1)c-\tfrac{i}{2}\,v^*_\mu(x)\,(\theta^\mu_1\gamma_2-\theta^\mu_2\gamma_1)\\[1mm]
&-\tfrac{i}{2}\,\xi^*_\mu(x)\,(\theta^\mu_1\gamma_2-\theta^\mu_2\gamma_1)c\;,
\end{split}    
\end{equation}
where we denoted $h:=h^\lambda_\lambda$ and $h^*:=h^{*\lambda}_\lambda\,$. 
It is possible to assign spacetime parity and ghost number to the component fields of \eqref{BVstringfield} by demanding the entire wave function $\Psi$ to have total even parity and ghost number zero. By doing so one can interpret $\Psi$ as a spacetime BV ``string field'', that contains the whole minimal BV spectrum plus auxiliary fields. In \eqref{BVstringfield} we have named the component fields accordingly: $h_{\mu\nu}$ and $h$ are the graviton and its trace, $v_\mu$ is an auxiliary vector and $\xi_\mu$ is the diffeomorphism ghost, the remaining components being all the corresponding anti-fields.

The BRST closure equation ${Q\Psi=0}$ at ghost number zero gives
\begin{equation}
\Box h_{\mu\nu}-\de_{(\mu}v_{\nu)}=0\;,\quad v_\mu+\de_\mu h-2\,\de\cdot h_\mu=0    
\end{equation}
that, solving for the auxiliary vector, yields the free spin two field equation
\begin{equation}
\Box h_{\mu\nu}-2\,\de_{(\mu}\de\cdot h_{\nu)}+\de_\mu\de_\nu h =0 \;.   
\end{equation}
The gauge symmetry $\delta h_{\mu\nu}=\de_{(\mu}\varepsilon_{\nu)}$ is recovered from the ghost number zero component of $\delta\Psi=Q\Lambda\,$, where
\begin{equation}\label{Lambdaparameter}
\Lambda=\varepsilon_\mu(x)\,(\theta^\mu_1\beta_2-\theta^\mu_2\beta_1)+\cdots    \;.
\end{equation}
The gauge parameter $\varepsilon_\mu$ should not be confused with the ghost $\xi_\mu$ appearing in \eqref{BVstringfield}! Indeed, in the string field interpretation of $\Psi\,$, the entire $\Lambda$ should have overall odd parity and ghost number $-1\,$, implying from \eqref{Lambdaparameter} that $\varepsilon_\mu$ has indeed even parity and ghost number zero, while $\xi_\mu$ has odd parity and ghost number $+1\,$. The \emph{spacetime} BRST transformations (indeed not to be confused with gauge symmetries) can also be obtained from the first-quantized BRST charge $Q$ as $s\Psi=Q\Psi\,$, where $s$ denotes the second-quantized BRST differential, in fact giving $s\,h_{\mu\nu}=\de_{(\mu}\xi_{\nu)}\,$.

\section{${\cal N}=4$ point particle in curved background}\label{curved}

In this section we are going to couple the model to a background metric $g_{\mu\nu}(x)\,$ taking the cohomological system \eqref{BRSTsystem} as a starting point for the deformation. In order to avoid $x$-dependent anticommutators in the fermionic sector, we define fermions to carry flat Lorentz indices, \emph{i.e.}  $(\theta_i^a, \bar\theta^{i\,a})$ and we introduce a background vielbein $e_\mu^a(x)$ and torsion-free spin connection\footnote{It is possible to avoid the introduction of background vielbein and spin connection, at the price of field-dependent hermiticity relations. For details see appendix \ref{flatvscurved}.} $\omega_{\mu\,ab}\,$.
Covariant derivative operators
\begin{equation}
\hat\nabla_\mu:=\de_\mu+\omega_{\mu\, ab}\,\theta^a\!\cdot\bar\theta^b\;,\quad\text{obey}\quad [\hat\nabla_\mu, \hat\nabla_\nu]=R_{\mu\nu\lambda\sigma}\,\theta^\lambda\!\cdot\bar\theta^\sigma=:\hat R_{\mu\nu}^\# \;, \end{equation}
where any fermion carrying a base vector index is understood as $\theta^\mu_i:=e^\mu_a(x)\,\theta^a_i\,$, same for $\bar\theta^{i\,\mu}\,$. 
 The curved space supercharges and laplacian are defined as\footnote{Note that $\nabla^2$ acts as the geometric laplacian, the second term in its definition has to be added to correctly rotate the index of the rightmost $\hat\nabla\,$.}
\begin{equation}
q_i:=-i\,\theta_i^a\,e^\mu_a\,\hat\nabla_\mu\;,\quad \bar q^i:=-i\,\bar\theta^{i\,a}\,e^\mu_a\,\hat\nabla_\mu\;,\quad \nabla^2:=g^{\mu\nu}\hat\nabla_\mu \hat\nabla_\nu-g^{\mu\nu}\,\Gamma^\lambda_{\mu\nu}\,\hat\nabla_\lambda    
\end{equation}
and obey
\begin{equation}
\begin{split}
&\{q_i, q_j\}=-\theta^\mu_i\theta^\nu_j\,\hat R_{\mu\nu}^\# \;,\quad \{\bar q^i, \bar q^j\}=-\bar\theta^{\mu\,i}\bar\theta^{\nu\,j}\,\hat R_{\mu\nu}^\#\;,\quad \{q_i, \bar q^j\}=-\delta_i^j\,\nabla^2-\theta^\mu_i\bar\theta^{\nu\,j}\,\hat R_{\mu\nu}^\#\;,\\[3mm]
& [\nabla^2, q_i]=i\,\theta^\mu_i\big(2\,\hat R_{\mu\nu}^\#\,\hat\nabla^\nu-\nabla^\lambda\hat R_{\lambda\mu}^\#-R_{\mu\nu}\hat\nabla^\nu\big)\;,\\[3mm]
& [\nabla^2, \bar q^i]=i\,\bar\theta^{\mu\,i}\big(2\,\hat R_{\mu\nu}^\#\,\hat\nabla^\nu-\nabla^\lambda\hat R_{\lambda\mu}^\#-R_{\mu\nu}\hat\nabla^\nu\big)\;.
\end{split}    
\end{equation}
In order to make an ansatz for the deformed BRST operator we assume that
\begin{itemize}
\item[i)] it has manifest background diffeomorphism invariance. In particular, spacetime derivatives are deformed only via minimal coupling (this rules out higher derivative couplings in the hamiltonian constraint).
\item[ii)] the ghost structure is the same as in the free theory (in particular we do not want to consider higher powers of ghost momenta),
\item[iii)] the non-minimal couplings to curvature in the hamiltonian have at most four fermions. This is due to the fact that the states in the reduced Hilbert space have at most spin two, making higher order fermionic couplings irrelevant.
\end{itemize}
Furthermore, consistency of the system \eqref{BRSTsystem} requires $[Q,{\cal T}_\alpha]=0\,$, at least weakly. This, together with the above assumptions, fixes the most general ansatz to be \begin{equation}\label{curvedQ}
\begin{split}
& Q:=c\,\D+\bm{\nabla}+\bar\gamma^i\gamma_i\,b\;,\quad\text{where}\\[2mm]
&\D:=\nabla^2+\alpha\,\hat R^{\#\#}+\Delta R\;,\quad \hat R^{\#\#}:=R_{\mu\nu\lambda\sigma}\,\theta^\mu\!\cdot\bar\theta^\nu\,\theta^\lambda\!\cdot\bar\theta^\sigma\;,\\[2mm]
& \bm{\nabla}:=\gamma_i\bar q^i+\bar\gamma^iq_i=-i\,S^\mu\hat\nabla_\mu\;,\quad S^\mu:=\bar\gamma^i\theta^\mu_i+\gamma_i\,\bar\theta^{\mu\,i}\;.
\end{split}    
\end{equation}
For brevity we grouped in $\Delta R$ all the couplings involving traces of the Riemann tensor that, according to the above assumptions, take the form
\begin{equation}
\Delta R = c_1\,R_{ab}\,S^a{}_c S^{cb}+c_2\,R\,S^{ab}S_{ab}+c_3\,R\;,    
\end{equation}
with $S^{ab}:=2\theta^{[a}\!\cdot\bar\theta^{b]}$ being the Lorentz generators.
The above parameters will be fixed to ensure nilpotency of $Q$ on suitable backgrounds:
\begin{equation}
Q^2=\bm{\nabla}^2+\bar\gamma\!\cdot\gamma\,\D+c\,[\D, \bm{\nabla}]\;.   
\end{equation}
The two obstructions are linearly independent, hence one should demand both $\bm{\nabla}^2+\bar\gamma\!\cdot\gamma\,\D=0$ and $[\D, \bm{\nabla}]=0\,$. The first term is explicitly given by
\begin{equation}\label{first obstruction}
 \bm{\nabla}^2+\bar\gamma\!\cdot\gamma\,\D = -\tfrac12\,S^\mu S^\nu\,\hat R_{\mu\nu}^\#+\bar\gamma\!\cdot\gamma\,(\alpha\,\hat R^{\#\#}+\Delta R)\;,
\end{equation}
and one can already see that $Q^2$ is obstructed, on the full Hilbert space, on any interesting background. However, the BRST cohomology describing the free graviton is defined on the reduced Hilbert space $\H_{\rm red}=\ker{\cal T}_\alpha\,$.
In order to evaluate the above expression on $\ker{\cal T}_\alpha\,$ we recall, as a preliminary step, that an arbitrary state in ${{\rm ker}\J^i_i}$ has the form
$$
\Phi_{AB}(x)\,Z^A_1Z^B_2+\chi_{AB}(x)\,Z^A_1Z^B_2\,c\quad\text{for}\quad Z_i^A=(\theta_i^a,\gamma_i,\beta_i)\;,
$$
and is thus annihilated by any obstruction of the form $O^{ABC}_{ijk}\bar Z_A^i\bar Z_B^j\bar Z_C^k\,$ where $O^{ABC}_{ijk}$ are arbitrary operators.
The expression \eqref{first obstruction} then becomes
\begin{equation}\label{first obstruction massage1}
    \bm{\nabla}^2+\bar\gamma\cdot\gamma\,\D\stackrel{{\rm ker}\J^i_i}{=} \bar\gamma\cdot\theta^\mu\,\gamma\cdot\bar\theta^\nu\,R_{\mu\nu}-\bar\gamma\cdot\gamma\,\left(\alpha\,R_{\mu\nu}\,\theta^\mu\!\cdot\bar\theta^\nu-\Delta R\rvert_{\ker\J_i^i}\right)\;.
\end{equation}
To proceed, we notice that the Young condition\footnote{The trace condition does not constrain the dependence on $Z^A_i\,$.} further constrains the $Z^A_i$ dependence of the states to
$$
\tfrac12\,\Phi_{AB}(x)\,Z^A_iZ^B_j\,\epsilon^{ij}+\tfrac12\,\chi_{AB}(x)\,Z^A_iZ^B_j\,\epsilon^{ij}\,c\;,
$$
as can be explicitly seen from \eqref{BVstringfield},
with $\epsilon^{ij}$ being the $su(2)$ antisymmetric tensor. Consequently, one has 
$$
\bar\gamma\cdot\theta^\mu\gamma\cdot\bar\theta^\nu+\bar\gamma\cdot\gamma\,\theta^\mu\!\cdot\bar\theta^\nu\stackrel{{\rm ker}{\cal T}\alpha}{=}0\;,
$$ 
that gives
\begin{equation}
 \bm{\nabla}^2+\bar\gamma\!\cdot\gamma\,\D\stackrel{{\rm ker}{\cal T}\alpha}{=}-\,\bar\gamma\cdot\gamma\left((\alpha+1)\,R_{\mu\nu}\,\theta^\mu\!\cdot\bar\theta^\nu-\Delta R\rvert_{\ker{\cal T}_\alpha}\right)  \;.  
\end{equation}
The second obstruction, evaluated on $\ker{\cal T}_\alpha$ yields
\begin{equation}\label{second obstruction}
\begin{split}
[\D, \bm{\nabla}] &= 2i(1-\alpha)\,S^\mu\hat R_{\mu\nu}^\#\hat\nabla^\nu-iS^\mu\nabla^\lambda\hat R_{\lambda
\mu}^\#+i(\alpha-1)S^\mu R_{\mu\nu}\hat\nabla^\nu+i\alpha S^\mu\nabla_\mu\hat R^{\#\#}+[\Delta R,\bm{\nabla}]\\[2mm]
& \!\!\!\stackrel{\ker{\cal T}_\alpha}{=} 2i(1-\alpha)\,S^\mu\hat R_{\mu\nu}^\#\hat\nabla^\nu-iS^\mu\nabla^\lambda\hat R_{\lambda
\mu}^\#+i(\alpha-1)S^\mu R_{\mu\nu}\hat\nabla^\nu\\
&+i\alpha \big(2 \nabla^\lambda\hat R_{\lambda
\mu}^\#\gamma\!\cdot\bar\theta^\mu-S^\mu\nabla_\mu R_{\nu\lambda}\theta^\nu\!\cdot\bar\theta^\lambda\big)+[\Delta R,\bm{\nabla}]\rvert_{\ker{\cal T}_\alpha}\;.
\end{split}    
\end{equation}
The first term above, ${2i(1-\alpha)S^\mu R_{\mu\nu\lambda\sigma}\theta^\lambda\!\cdot\bar\theta^\sigma}\hat\nabla^\nu\,$, involves the full Riemann tensor, thus fixing $\alpha=1\,$. Similar terms, namely with the derivative acting through to the right, are produced by $[\Delta R,\bm{\nabla}]\rvert_{\ker{\cal T}_\alpha}$ and cannot be canceled by other terms. This prevents having fermions in $\Delta R\,$, fixing $c_1=c_2=0$ and giving
\begin{equation}
\begin{split}
Q^2 &\stackrel{\ker{\cal T}_\alpha}{=} \bar\gamma\cdot\gamma\left(c_3\, R-2\,R_{\mu\nu}\,\theta^\mu\!\cdot\bar\theta^\nu\right)\\[2mm]
&+c\left\{4i\,\nabla_{[\nu}R_{\lambda]\mu}\,\theta^\nu\!\cdot\bar\theta^\lambda\gamma\!\cdot\bar\theta^\mu-i(\nabla_\mu R_{\nu\lambda}+2\nabla_{[\lambda} R_{\nu]\mu})S^\mu\theta^\nu\!\cdot\bar\theta^\lambda+ic_3\,S^\mu\nabla_\mu R\right\}\;. 
\end{split}
\end{equation}
We can thus achieve nilpotency of the BRST charge only on Einstein manifolds, \emph{i.e.} obeying $R_{\mu\nu}=\lambda\,g_{\mu\nu}$
by choosing $c_3=\tfrac{2}{d}\,$:
\begin{equation}
Q^2 \stackrel{\ker{\cal T_\alpha}}{=}0\quad\text{for}\quad R_{\mu\nu}=\lambda\,g_{\mu\nu}\;,\quad\D=\nabla^2+\hat R^{\#\#}+2\lambda\;.  
\end{equation}
We would like to remark that nilpotency of the BRST operator, that is quantum consistency of the worldline system, determines Einstein equations for the background in contrast to string theory where world sheet conformal invariance implies Ricci flatness (modulo $\alpha'$ corrections). The worldline theory thus reproduces what is expected form General Relativity.

In order to display the field equations for the graviton in curved background, one repeats the same analysis of the previous section by using the deformed BRST charge \eqref{curvedQ} with the appropriate choice $\D=\nabla^2+\hat R^{\#\#}+2\lambda\,$, yielding
\begin{equation}
\nabla^2h_{\mu\nu}-2\nabla_{(\mu}\nabla\cdot h_{\nu)}+\nabla_\mu\nabla_\nu h^\lambda_\lambda+2W_{\mu\lambda\nu\sigma}\,h^{\lambda\sigma}+\tfrac{2\lambda}{d-1}\,\big(g_{\mu\nu}\,h^\lambda_\lambda-h_{\mu\nu}\big)=0\;,    
\end{equation}
where $W_{\mu\nu\lambda\sigma}$ is the traceless Weyl tensor.
One can verify that the above equation corresponds to the linearization of $R_{\mu\nu}(g+h)=\lambda(g_{\mu\nu}+h_{\mu\nu})$ around an Einstein background $g_{\mu\nu}\,$, thus confirming that the spin two self couplings coincide with those of Einstein's gravity.

\section{Vertex operators and three graviton amplitude}

In the worldline formalism, one usually derives vertex operators starting from an interacting lagrangian $\cal L\,$, that is expanded around its free part ${\cal L}_0$ in powers of the background fields fluctuations. The (linear) vertex operator $W_0$ is then defined as
\begin{equation}
W_0:=({\cal L}-{\cal L}_0)\rvert_{\text{linear in BG fields}}\;.    
\end{equation}
In general non-linear vertices, defined by higher order terms in $({\cal L}-{\cal L}_0)\,$, are also needed, in order to take care of diagrams with multiple legs joining at a single point (pinching vertices). In our present setting, that is hamiltonian BRST, one can define the interacting gauge fixed hamiltonian by ${H:=\{Q, b\}}\,$, from which one can derive the vertex operator. This is sufficient to compute one-particle irreducible diagrams at one-loop, since in that case the worldline is naturally associated to the loop, from which external states stick out as vertex operators. Treating tree-level amplitudes, however, is in general more complicated. Especially for self-interacting theories like pure Yang-Mills and gravity, all internal and external lines are of the same species, and there is no natural worldline to be found. For a given tree-level diagram one has thus to \emph{choose} an appropriate line connecting two external states to be the worldline. By doing so one is selecting two external states to be at the (asymptotic) endpoints of the worldline, to be specified by the boundary conditions of the path integral, while the other external states are carried by the vertex operators discussed above. 
The corresponding diagram is thus given by the expectation value
\begin{equation}\label{naiveamplitude}
\bra{f}T\{W_{0,1}(\tau_1)\prod_{k=2}^{n-2}\int d\tau_k\,W_{0,k}(\tau_k)\}
\ket{i}    
\end{equation}
where $\ket{i}$ and $\ket{f}$ label the two external states that are chosen to be at the worldline endpoints. As displayed above, vertex operators in general have to be integrated over the worldline. This is easily understood, as they come from deformations of the action. In other terms, more akin to string theory, their positions are moduli that must be integrated over. One vertex operator (the $W_{0,1}(\tau_1)$ above), however, has to be fixed to an arbitrary position in order to fix the leftover invariance under rigid translations of the worldline. This is the point particle analogue of fixing three vertex operators in string theory at tree-level, to remove the conformal Killing symmetries left after gauge fixing. In BRST Hilbert space language, this is equivalent to the fact that the inner product between conformal vacua needs three ghost insertions to be nonzero: $\bra{0}c_{-1}c_0c_1\ket{0}\sim 1\,$ whose point particle analogue is in fact $\bra{0}c\ket{0}\sim1\,$. One obvious difference is that in string theory conformal invariance implies an  operator-state correspondence so that all the external states can be treated on equal footing as vertex operators attached to a vacuum worldsheet. In the RNS superstring there is an extra subtlety in that 
two of the three unintegrated vertex operators should be in  picture $-1\,$, 
while the third unintegrated vertex and the integrated ones should be in picture zero. This suggests that a similar mechanism may be at work for the point particle \cite{Dai:2008bh}, such that the initial and final states in \eqref{naiveamplitude} can be obtained by vertex operators acting on the vacuum, such that $\bra{f}...\ket{i}=\bra{0}V_f...V_i\ket{0}\,$. To explore this idea for the model at hand\footnote{In \cite{Dai:2008bh} this analysis was performed for the case of ${\cal N}=2$ spinning particle in a Yang-Mills background, analogous to the open string.} we shall return to the hamiltonian BRST treatment.

Following \cite{Berkovits:1991gj,Dai:2008bh,SiegelRNS}, let us consider a first-quantized system with interacting\footnote{Typically this describes a point particle interacting with background spacetime fields.} BRST charge $Q\,$, and expand it around its free part as
\begin{equation}
Q=Q_0+V\;.  
\end{equation}
Nilpotency of the full BRST charge yields for the vertex
\begin{equation}
\{Q_0, V\}+V^2=0\;,    
\end{equation}
while the part of $V$ linear in fluctuations, that we denote $V_0\,$, is closed with respect to the free BRST charge: $\{Q_0, V_0\}=0\,$. Suppose now that the BRST Hilbert space contains a vacuum $\ket{\tilde 0}\,$, usually different from the Fock vacuum $\ket{0}$ for ghosts, that is also a physical state\footnote{For the case at hand this entails, besides being in the cohomology of $Q_0\,$, to be in the kernel of ${\cal T_\alpha}\,$.} and has ghost number $-1$.
Acting with $V_0$ on this vacuum will thus produce a physical state at ghost number zero:
\begin{equation}\label{correspondence}
\ket{V}:=V_0\ket{\tilde0}\quad\Rightarrow\quad Q_0\ket{V}=0\;,\quad \delta\ket{V}=Q_0\ket{\Lambda}\;,    
\end{equation}
the gauge invariance descending from $V_0\sim V_0+[Q_0,\Lambda]$ with $\Lambda$ a ghost number minus one operator parameter. 
The correspondence sketched by \eqref{correspondence} works directly for the case of the ${\N=2}$ spinning particle on a Yang-Mills background: the ``physical state'' vacuum is $\ket{\tilde0}:=\beta\ket{0}\,$ where $\ket{0}$ is the Fock vacuum and $\beta$ the first-quantized antighost creator\footnote{The BRST Hilbert space for ${\N=2}$ coincides with a single sector of the ${\N=4}$ case considered here, \emph{i.e.} ${(\theta^\mu_i, \gamma_i, \beta_i)\to(\theta^\mu, \gamma, \beta)}\,$, same for the barred operators.}. In the string field interpretation of the ${\N=2}$ BRST Hilbert space, $\ket{\tilde0}$ corresponds to a constant Yang-Mills ghost. The corresponding linear vertex operator, defined from $Q=Q_0+V_0(A)+{\cal O}(A^2)\,$ reads (see \cite{Dai:2008bh} for details) 
\begin{equation}\label{YMvertex}
V_0= c\,\big[2\,A\cdot p-4i\,\de_\mu A_\nu\,\theta^{[\mu}\bar\theta^{\nu]}\big]-(\gamma\bar\theta^\mu+\bar\gamma\theta^\mu)\,A_\mu=:c\,W_{\text{I}}+W_{\text{II}}  \end{equation}
in background Lorentz gauge $\de\cdot A=0\,$. When acting on the vacuum $\ket{\tilde0}$ it produces the physical vector state:
$V_0\ket{\tilde0}=A_\mu\,\theta^\mu\ket{0}$ while containing information about the integrated vertex as well: The $W_{\text{II}}$ part of the vertex, that creates the vector from the YM ghost vacuum, corresponds to the picture $-1$ vertex of the open string\footnote{The difference in the ghost dependence with respect to open string theory depends both on the choice of vacuum $\ket{0}$ and in the smaller Killing group of the point particle.} with $W_{\text{I}}$ corresponding to the picture zero.

For ${\N=4}$ the situation is more complicated. Let us consider the expansion of the background vielbein around flat space as $e^a_\mu = \delta^a_\mu +\tilde{e}^a_\mu\,$. Since the introduction of the vielbein itself is a mere technical point (see appendix \ref{flatvscurved} for details), we can choose a local Lorentz frame where the antisymmetric part of the fluctuation vanishes: $\tilde e_{[\mu\nu]}:=\delta^a{}_{[\mu}\tilde e_{\nu]a}=0$ yielding $h_{\mu\nu}:=2\,\delta^a{}_{(\mu}\tilde e_{\nu)a}=2\,\tilde e_{\mu\nu}$ for the metric fluctuation or, equivalently, $e^a_\mu=\delta^a_\mu+\tfrac12\,h^a_\mu\,$, where now we switch between flat and curved indices with $\delta^a_\mu\,$.
The linear vertex operator 
\begin{equation}
\begin{split}
V_0=\left(Q-Q_0\right)_{\text{linear in }h \text{ and }\tilde e}\;,
\end{split}
\end{equation}
resulting from the expansion of \eqref{curvedQ} reads
\begin{equation} \label{VertexO}
\begin{split}
{V_0} &= c\left( -{ {h^{\mu \nu }}{\de_\mu }{\de_\nu } + \left( {{\omega _{\mu ab}}\,{\de^\mu } + {\de^\mu }\,{\omega _{\mu ab}}} \right)\theta^a\!\cdot\bar\theta^b -{\eta ^{\mu \sigma }}\Gamma _{\mu \sigma }^\nu {\de_\nu } + {R_{abcd}}\theta^a\!\cdot\bar \theta^b\,\theta^c\!\cdot \bar \theta ^d} \right) \\
 &+i \left( {\left( {{{\bar \gamma }}\!\cdot\theta^c + {\gamma }\!\cdot\bar \theta ^c} \right)\tilde e_c^\mu {\de_\mu } -\left( {{{\bar \gamma }}\!\cdot\theta^\mu  + {\gamma}\!\cdot\bar \theta ^\mu } \right)\theta^a\!\cdot\bar\theta^b{\omega _{\mu ab}}} \right) \\
 & = c\left( {  {h^{\mu \nu }}{p_\mu }{p_\nu } - 2i\,{\partial _{[\nu }}{h_{\lambda ]\mu }}\,{p^\mu }\theta^\nu\!\cdot \bar \theta^\lambda  - 2\left( {{\partial _{[\lambda }}{\partial _{[\mu }}{h_{\nu ]\sigma ]}}} \right)\theta^\mu\!\cdot \bar \theta^\nu\, \theta^\lambda\!\cdot \bar \theta^\sigma } \right)\\
 &- \tfrac{1}{2}\left( {{{\bar \gamma }}\!\cdot\theta^\mu  + {\gamma}\!\cdot\bar \theta^\mu } \right){h_{\mu \nu }}\,{p^\nu } +i \left( {{{\bar \gamma }}\!\cdot\theta^\mu  + {\gamma}\!\cdot\bar \theta ^\mu } \right){\partial _{[\nu }}{h_{\lambda ]\mu }}\theta^\nu\!\cdot \bar \theta^\lambda =: c\,W_{\text{I}}+W_{\text{II}} \;,
\end{split}
\end{equation}
where we used $\partial^\mu h_{\mu\nu} =\eta^{\mu\nu}h_{\mu\nu} = 0\,$. We recall (see appendix \ref{flatvscurved}) that the relation between partial derivatives and momenta is given by $\de_\mu=i\,g^{1/4}p_\mu\,g^{-1/4}=i\,p_\mu-\tfrac14\,\de_\mu h^\lambda_\lambda+O(h^2)$ that reduces to $\de_\mu=i\,p_\mu$ at the linearized level for a traceless graviton, thus yielding \eqref{VertexO}.

Contrary to the Yang-Mills case, in the ${\N=4}$ Hilbert space $\H_{\rm red}$ there is no scalar physical vacuum.
Let us instead consider the vector state lying at ghost number $-1\,$, corresponding to the diffeomorphism ghost:
\begin{equation}\label{vectorghost}
\ket{\xi}:=\xi_\mu\,(\theta_1^\mu\beta_2-\theta_2^\mu\beta_1)\ket{0} \;, \end{equation}
that, in order to be in the cohomology of $Q_0\,$, has to be a Killing vector of flat spacetime: $\de_{(\mu}\xi_{\nu)}=0\,$.
We will now restrict the graviton fluctuation to be a plane wave: $h_{\mu\nu}=\varepsilon_{\mu\nu}\,e^{ik\cdot x}$ with $k\cdot\varepsilon_{\mu}=\varepsilon^\mu_\mu=0\,$. The vertex operator $V_0 = cW_{\text{I}} + W_{\text{II}}$ thus reduces to
\begin{equation}
\begin{split}
W_{\text{I}} &= \varepsilon_{\mu\nu}\left(p^\mu p^\nu - k_\sigma S^{\nu\sigma}p^\mu - \frac{1}{2}k_\sigma k_\rho S^{\sigma\mu}S^{\nu\rho}\right)e^{ik\cdot x}, \\
W_{\text{II}} &= -\tfrac{1}{2}\varepsilon_{\mu\nu}\left(S^\nu p^\mu +k_\sigma S^\mu S^{\sigma\nu}\right)e^{ik\cdot x},
\end{split}
\end{equation}
where we recall that $S^\mu = \bar{\gamma}\cdot \theta^\mu +\gamma\cdot \bar{\theta}^{ \mu}$ and $S^{\mu\nu} =2 \theta^{[\mu}\!\cdot\bar{\theta}^{\nu]}$.
For a given on-shell graviton state with polarization $\varepsilon_{\mu\nu}$ and momentum $k_\mu$ one can choose a constant vector $\xi_\mu$ obeying $\xi\cdot\varepsilon_\mu=0$ and $\xi\cdot k=1\,$. By acting on such state \eqref{vectorghost} with $W_{\text{II}}$ one obtains
\begin{equation}
\begin{split}
W_{\text{II}} \ket{\xi}&= \varepsilon_{\mu\nu}\left(\xi \cdot k\, \theta_1^\mu \theta_2^\nu - 2\xi^\nu k_\sigma \theta_{[1}^\sigma \theta_{2]}^\mu\right) e^{ik\cdot x}\left|0\right> \\
&= \varepsilon_{\mu\nu} \theta_1^\mu \theta_2^\nu e^{ik\cdot x}\left|0\right>=\ket{h}\;.
\end{split}
\end{equation}
It is thus possible to use the BRST deformation to obtain a graviton state out of its own ghost. Contrary to the Yang-Mills case, however, one cannot use the ghost state as a viable vacuum since it breaks Lorentz invariance and is not universal, being a different vector for each graviton. 

According to the above discussion, an $n$-graviton tree-level world line diagram is given by
\begin{equation}
\bra{h^{(1)}}T\{V^{(2)}\prod_{i=3}^{n-1}\int d\tau_i\,W_{\text{I}}^{(i)}\}\ket{h^{(n)}}=\bra{\xi^{(1)}}T\{V^{(1)}V^{(2)}\prod_{i=3}^{n-1}\int d\tau_i\,W_{\text{I}}^{(i)}V^{(n)}\}\ket{\xi^{(n)}}\;.
\end{equation}
Applying the above formula to the three point function one obtains
\begin{equation}
\begin{split}
\left\langle h^{(3)} \right|{V^{(2)}}\left| h^{(1)} \right\rangle  &= \bra{\xi^{(3)}}T\{V^{(3)}V^{(2)}V^{(1)}\}\ket{\xi^{(1)}}\\
 &= \left( \begin{array}{l}
   \text{Tr}\left( {{\varepsilon ^{\left( 2 \right)}} \cdot {\varepsilon ^{\left( 1 \right)}}} \right)\left( {{k_1} \cdot {\varepsilon ^{\left( 3 \right)}} \cdot {k_1}} \right) + \text{Tr}\left( {{\varepsilon ^{\left( 3 \right)}} \cdot {\varepsilon ^{\left( 2 \right)}}} \right)\left( {{k_2} \cdot {\varepsilon ^{\left( 1 \right)}} \cdot {k_2}} \right)\\ + \text{Tr}\left( {{\varepsilon ^{\left( 3 \right)}} \cdot {\varepsilon ^{\left( 1 \right)}}} \right)\left( {{k_3} \cdot {\varepsilon ^{\left( 2 \right)}} \cdot {k_3}} \right)  \\
 - 2\left( {{k_1} \cdot {\varepsilon ^{\left( 3 \right)}} \cdot {\varepsilon ^{\left( 1 \right)}} \cdot {\varepsilon ^{\left( 2 \right)}} \cdot {k_1} + {k_2} \cdot  { {\varepsilon ^{\left( 3 \right)}} \cdot {\varepsilon ^{\left( 2 \right)}} \cdot {\varepsilon ^{\left( 1 \right)}}}\cdot {k_2} + {k_3} \cdot {\varepsilon ^{\left( 1 \right)}} \cdot {\varepsilon ^{\left( 3 \right)}} \cdot {\varepsilon ^{\left( 2 \right)}} \cdot k_3} \right)
\end{array} \right) \\
& =  \varepsilon^{(1) \mu \alpha }\varepsilon^{(2) \nu \beta }\varepsilon^{(3) \sigma \gamma }\left( {{k_1}_\sigma {\eta _{\mu \nu }} + {k_2}_\mu {\eta _{\nu \sigma }} + {k_3}_\nu {\eta _{\sigma \mu }}} \right)\left( {{k_1}_\gamma {\eta _{\alpha \beta }} + {k_2}_\alpha {\eta _{\beta \gamma }} + {k_3}_\beta {\eta _{\gamma \alpha }}} \right).
\end{split}
\end{equation}
On-shell, the contractions of the exponentials in the plane waves give a factor of one. This is in agreement with the on-shell three-graviton vertex in Einstein gravity.
In order to develop an efficient formalism to compute tree-level graviton scattering with this model, it would be interesting to adapt the so-called worldgraph approach of \cite{Dai:2006vj,Dai:2008bh}, that should result in simpler ``Feynman rules'' compared to the standard worldline computations. 
As for the implementation of the constraints ${\cal T}_\alpha\ket{\Psi}=0$ in the path integral, it is not a problem for tree-level amplitudes: since the gauge fixed hamiltonian commutes with the constraints, a worldline with an asymptotic state in ${\cal H}_{\rm red}=\ker{\cal T}_\alpha$ will keep it in
${\cal H}_{\rm red}$ throughout its evolution. On the other hand, for one-loop amplitudes and higher the situation is different, as the worldline loop corresponds to a trace over the full Hilbert space. It is thus necessary to find the correct way to perform the projection on the Hilbert space, \emph{i.e.} construct the appropriate one-loop measure for the path integral, that should be related to the $R$-symmetry gauging used in \cite{Bastianelli:2007pv,Bastianelli:2012bn}.

\section{Conclusions}

It is well known that the ${\N=4}$ spinning particle describes free gravitons in Minkowski spacetime. Despite the obvious existence of non-linear self-interacting gravity, it was so far not known how to couple the ${\N=4}$ worldline action to a curved background, due to curvature obstructions that break the supersymmetry algebra. 

Inspired by the work of \cite{Dai:2008bh} for Yang-Mills, in this note we have coupled the ${\N=4}$ spinning particle to background gravity at the level of hamiltonian BRST. The coupling has manifest background diffeomorphism invariance at the full non-linear level. We then showed that quantum consistency of the worldline model, which amounts to nilpotency of the BRST operator, requires the gravitational background to satisfy Einstein's equations. Similarly to what happens in the case of ${\N=2}$ coupled to Yang-Mills \cite{Dai:2008bh}, it is necessary to truncate the ${\N=4}$ Fock space to have a consistent coupling. We also found that infinitesimal variations of the BRST operator around a classical solution produce an unintegrated, picture zero operator which generates an asymptotic graviton state when acting on a diffeomorphism ghost state. Furthermore, the simplest world graph with 3 vertex operators reproduces the 3-graviton coupling in General Relativity. As such, the $\N=4$ spinning particle serves as a useful toy model for a background independent formulation of string field theory. Indeed, given a manifestly background independent BRST quantization of the worldline this BRST charge can, in principle, be integrated to obtain the homological vector field of the space of fields which, in turn, can be taken as a starting point for the BV quantization. 

An interesting extension of the present work is to relax the constraint on the Hilbert space to subalgebras of $so(4)$ which should give rise to the complete massless sector of the NS-sector of string teory, including the Kalb-Ramond field $B_{\mu\nu}$ as well as the dilaton and their coupling to the background fields through deformations of the BRST charge that preserve a suitable subgroup of the $SO(4)$ $R$-symmetry. We will return to this problem in a forthcoming work.

\section*{Acknowledgments}
We would like to thank Maxim Grigoriev and Allison Pinto for discussions as well as Warren Siegel for helpful comments. R.B. wants to thank Nicolas Boulanger, David de Filippi, Fiorenzo Bastianelli, Olindo Corradini, Emanuele Latini for useful discussions and LMU University for warm hospitality during completion of this work. I.S. would like to thank University of Mons for hospitality where this work was initiated. The work of R.B. was supported by a PDR “Gravity and extensions” from the F.R.S.-FNRS (Belgium). A.M. and I.S. were supported, in parts, by the DFG Transregional Collaborative Research Centre TRR 33 and the DFG cluster of excellence "Origin and Structure of the Universe".

\appendix
\section{Flat vs curved fermions $\&$ hermiticity}\label{flatvscurved}

In worldline applications in curved space it is customary, when worldline RNS\footnote{Borrowing from string theory language, we simply mean worldline oscillators, either fermionic or bosonic, that are spacetime vectors.} oscillators are
present, to introduce spacetime background vielbein and spin connection, even though there are no spacetime fermions, and treat the oscillators as spacetime flat Lorentz vectors, just as it happens for spacetime gamma matrices in curved space. This is mostly to avoid $x$-dependent commutation relations between the oscillators. It is nonetheless viable to use curved base indices and avoid the introduction of background vielbeins. In the following we explicitly present the mapping between the two formalisms in the case of a particle model with ${{\cal N}=2s}$ supersymmetries.

\subsection{Flat fermions}

Here we consider a $d$-dimensional target spacetime manifold ${\cal M}$ endowed with a vielbein\footnote{The metric and the Christoffel symbols descend in the obvious way with $g_{\mu\nu}=e_\mu^a\,e_{\nu\,a}\,$.} $e_\mu^a(x)$ and torsion-free spin connection $\omega_{\mu\,ab}(x)\,$. The graded phase space has coordinates $(p_\mu, x^\mu, \theta^a_i, \bar\theta^{a\,i})$ with $i=1,...,{\cal N}/2$ fermionic oscillator families (${\cal N}=4$ for the model used in the paper). The symplectic current
\begin{equation}
\Theta:=p_\mu\dot x^\mu+i\,\bar\theta^i_a\dot\theta^a_i    
\end{equation}
induces the quantum (anti)-commutation relations
\begin{equation}
[x^\mu, p_\nu]=i\,\delta^\mu_\nu\;,\quad \{\theta^a_i, \bar\theta^{b\,j}\}=\delta^j_i\,\eta^{ab}\;,
\end{equation}
the other (anti)-commutators being zero. The states in the Hilbert space, that are subject to suitable physical state conditions, are isomorphic\footnote{One can consider the usual Fock space realization of the fermionic algebra, by choosing a vacuum $\ket{0}$ obeying $\bar\theta^{a\,i}\ket{0}=0$ and identify (hiding all indices) $\psi_{...}(x)\theta...\theta\ket{0}\sim\psi(x, \theta)$} to wave functions $\psi(x, \theta_i)$ that consist of spacetime multiforms, as explicitly displayed in \eqref{wavefunction} for the present case of ${\cal N}=4\,$. In the following discussion the number of fermion families is immaterial and we often use a vector state (pertinent to ${\cal N}=2$) $\ket{A}\sim A_a\,\theta^a\equiv A_\mu\,e^\mu_a\,\theta^a$ for examples.
The covariant inner product is defined by
\begin{equation}
\braket{V}{W}:=\int d^dx\sqrt{g}\,V^*_a\,W^a
\end{equation}
for vectors, which generalizes to multi-forms\footnote{We remind that a multi index $a[n]$ stands for $[a_1...a_n]$ antisymmetrized with strength one.}
\begin{equation}\label{flatbraket}
\braket{\chi}{\phi}:=c_{s,n}\,\int d^dx\sqrt{g}\,\chi^*_{a_1[n_1]...a_s[n_s]}\,\phi^{a_1[n_1]...a_s[n_s]}\;.
\end{equation}
In the fermionic Hilbert space this coincides with the Fock inner product, giving $(\theta^a_i)^\dagger=\bar\theta^{a\,i}\,$ while the metric determinant in \eqref{flatbraket} yields the identification
\begin{equation}
g^{1/4}p_\mu\,g^{-1/4}=-i\de_\mu    
\end{equation}
for a self-adjoint momentum operator $p_\mu^\dagger=p_\mu\,$.
Covariant momenta and derivative operators are defined as
\begin{equation}
\pi_\mu:=p_\mu-i\,\omega_{\mu\, ab}\,\theta^a\!\cdot\bar\theta^b\;,\quad \hat\nabla_\mu:=i\,g^{1/4}\pi_\mu\,g^{-1/4}=\de_\mu+\omega_{\mu\, ab}\,\theta^a\!\cdot\bar\theta^b    
\end{equation}
and obey $\pi_\mu^\dagger=\pi_\mu\,$, $\hat\nabla_\mu^\dagger=-(\hat\nabla_\mu+\Gamma^\lambda_{\mu\lambda})$ with respect to the above inner product. The supercharges
\begin{equation}
q_i:=-i\theta^a_i\,e^\mu_a\,\hat\nabla_\mu\;,\quad \bar q^i:=-i\bar\theta^{a\,i}\,e^\mu_a\,\hat\nabla_\mu    
\end{equation}
are related by the adjoint operation: $(q_i)^\dagger=\bar q^i$ and the self-adjoint covariant laplacian reads
\begin{equation}
\nabla^2:=g^{\mu\nu}\hat\nabla_\mu\hat\nabla_\nu-g^{\mu\nu}\Gamma^\lambda_{\mu\nu}\hat\nabla_\lambda\equiv \frac{1}{\sqrt{g}}\hat\nabla_\mu\,g^{\mu\nu}\sqrt{g}\,\hat\nabla_\nu\;.    
\end{equation}

\subsection{Curved fermions}

In this case we will only introduce a metric $g_{\mu\nu}(x)$ on the target space manifold, together with its Levi-Civita connection. The coordinates of the graded phase space are chosen as $(p_\mu, x^\mu, \theta^\mu_i, \bar\theta^i_\mu)\,$. The symplectic current
\begin{equation}
\Theta:=p_\mu\dot x^\mu+i\,\bar\theta^i_\mu\dot\theta^\mu_i   \end{equation}
gives the non vanishing (anti)-commutators
\begin{equation}
[x^\mu, p_\nu]=i\,\delta^\mu_\nu\;,\quad \{\theta^\mu_i, \bar\theta^j_\nu\}=\delta^j_i\,\delta^\mu_\nu\;,
\end{equation}
where the independent fermionic momentum is the covector $\bar\theta^i_\mu\;,$ while  ${\bar\theta^{\mu\,i}:=g^{\mu\nu}(x)\,\bar\theta^i_\nu}\,$. Tensors in the wave functions now carry base curved indices, \emph{e.g.} $\ket{A}\sim A_\mu\,\theta^\mu$
and the natural inner product becomes
\begin{equation}\label{curvedbraket}
\braket{V}{W}:=\int d^dx\sqrt{g}\,V^{*\mu}\,W_\mu\equiv \int d^dx\sqrt{g}\,g^{\mu\nu}V^*_\mu\,W_\nu\;,
\end{equation}
with obvious generalization to multi-forms.
The adjoint operation on fermions now involves the spacetime metric: \begin{equation}\label{thetadagger}
(\theta^\mu_i)^\dagger=g^{\mu\nu}(x)\bar\theta^i_\nu\;,
\end{equation}
consistently with \eqref{curvedbraket}. The partial derivative is again related to the momentum operator by
$g^{1/4}p_\mu\,g^{-1/4}=-i\de_\mu$ but, by taking the adjoint of the relation $[p_\mu, \theta^\nu_i]=0\,$, consistency with \eqref{thetadagger} implies that
\begin{equation}\label{pdagger}
p_\mu^\dagger=p_\mu+i\,\de_\mu g_{\nu\lambda}\,\theta^\nu\!\cdot\bar\theta^\lambda\;,    
\end{equation}
as it can also be deduced from \eqref{curvedbraket}.
The covariant momenta, derivatives and supercharges are defined by
\begin{equation}
\begin{split}
&\pi_\mu:=p_\mu+i\,\Gamma^\lambda_{\mu\nu}\,\theta^\nu\!\cdot\bar\theta_\lambda\;,\quad \hat\nabla_\mu:=i\,g^{1/4}\pi_\mu\,g^{-1/4}=\de_\mu-\Gamma^\lambda_{\mu\nu}\,\theta^\nu\!\cdot\bar\theta_\lambda\;,\\[2mm]
&q_i:=-i\theta^\mu_i\,\hat\nabla_\mu\;,\quad \bar q^i:=-i\bar\theta^i_\mu\,g^{\mu\nu}\,\hat\nabla_\nu 
\end{split}
\end{equation}
and still obey $\pi_\mu^\dagger=\pi_\mu\,$, $\hat\nabla_\mu^\dagger=-(\hat\nabla_\mu+\Gamma^\lambda_{\mu\lambda})$ and $(q_i)^\dagger=\bar q^i$ with respect to \eqref{curvedbraket}. Correspondingly, the covariant laplacian 
\begin{equation}
\nabla^2:=g^{\mu\nu}\hat\nabla_\mu\hat\nabla_\nu-g^{\mu\nu}\Gamma^\lambda_{\mu\nu}\hat\nabla_\lambda\equiv \frac{1}{\sqrt{g}}\hat\nabla_\mu\,g^{\mu\nu}\sqrt{g}\,\hat\nabla_\nu    
\end{equation}
is self adjoint.

\paragraph{Mapping the  two}

The map between the two realizations starts from the obvious redefinition of the fermionic oscillators\footnote{Notice that, in a path integral formulation, the Jacobians cancel from the measure and $D\bar\theta_a D\theta^a\equiv D\bar\theta_\mu D\theta^\mu$}
\begin{equation}\label{thetamap}
\theta^\mu_i=e^\mu_a(x)\,\theta^a_i\;,\quad \bar\theta_\mu^i=e_{\mu\,a}(x)\,\bar\theta^{a\,i}\;.    
\end{equation}
The transformation between momenta can be found by the requirement
\begin{equation}
[p_\mu^{\rm flat}, \theta^a_i]=0   \;,\quad [p_\mu^{\rm curved}, \theta^\nu_i]=0 
\end{equation}
provided \eqref{thetamap}. This fixes
\begin{equation}
p_\mu^{\rm curved}=p_\mu^{\rm flat}-i\,\de_\mu e_{\nu\,a}\,\theta^\nu\!\cdot\bar\theta^a\;, \end{equation}
that is consistent with the hermiticity properties displayed above and preserves the symplectic current:
\begin{equation}
\Theta:=p^{\rm curved}_\mu\dot x^\mu+i\,\bar\theta^i_\mu\dot\theta^\mu_i=p^{\rm flat}_\mu\dot x^\mu+i\,\bar\theta^i_a\dot\theta^a_i\;,   \end{equation}
thus providing a canonical transformation in the graded phase space.
Given the above redefinitions, covariant momenta and derivatives coincide, namely $\pi_\mu^{\rm curved}=\pi_\mu^{\rm flat}\,$, $\hat\nabla_\mu^{\rm curved}=\hat\nabla_\mu^{\rm flat}\,$ and hence so do the supercharges and covariant laplacian.

\section{$so(4)$ algebra}

In this subsection we provide a detailed calculation of the algebra of $so(4)$ $R$-symmetry generators. We define the full (physical fields and ghost fields) $so(4)$ generators in the following way,
\begin{equation}
\begin{split}
\J_{IJ}= i\,\Theta^\mu_{[I}\Theta_{J]\mu} - 2i\,\mathcal{B}_{[I}\Gamma_{J]},
\end{split}
\end{equation}
where $I \in \{1,2,3,4\}\,$. The $\Theta_I$ are fermionic fields which obey the anticommutation relation $\{\Theta^\mu_I,\Theta_{J\nu}\} = \delta_{IJ}\delta^\mu_\nu$, the $\mathcal{B}_I$ and  $\Gamma_I$ are bosonic ghosts satisfying the commutation relation, ${[\mathcal{B}_I,\Gamma_J] = 2\delta_{IJ}}$. The $so(4)$ generators satisfy the following commutation relation:
\begin{equation}
\begin{split}
[\J_{IJ},\J_{KL}]=i\left(\delta_{JK}\J_{IL}-\delta_{JL}\J_{IK}-\delta_{IK}\J_{JL}+\delta_{IL}\J_{JK}\right).
\end{split}
\end{equation}
It is more convenient to work with a complex basis by definig
\begin{equation}
\begin{split}
&\theta^\mu_i = \frac{1}{\sqrt{2}}\left(\Theta^\mu_i + i \Theta^\mu_{i+2}\right), \ \ \bar{\theta}^i_\mu = \frac{1}{\sqrt{2}}\left(\Theta_{\mu i} - i \Theta_{\mu i+2}\right), \\
&\beta_i = \frac{1}{2}\left(\mathcal{B}_i + i \mathcal{B}_{i+2}\right), \ \ \ \ \ \bar{\beta}^i = \frac{1}{2}\left(\mathcal{B}_i - i \mathcal{B}_{i+2}\right), \\
&\gamma_i = \frac{1}{2}\left(\Gamma_i + i \Gamma_{i+2}\right), \ \ \ \ \ \bar{\gamma}^i = \frac{1}{2}\left(\Gamma_i - i \Gamma_{i+2}\right),
\end{split}
\end{equation}
where $i \in \{1,2\}$. The commutation relations in this basis are,
\begin{equation}
\begin{split}
\{\theta^\mu_i,\bar{\theta}^j_\nu\} = \delta^\mu_\nu\delta^j_i, \ \ \ [\beta_i,\bar{\gamma}^j] = [\bar{\beta}^j,\gamma_i] = \delta^j_i, 
\end{split}
\end{equation}
where all the other commutation relations vanish.
We define a new set of generators built out from the old $\J_{IJ}$ generators in the following way,
\begin{equation}
\begin{split}
&{\J^1_1} \equiv -{\J_{13}} = {\theta _1}{{\bar \theta }^1} + {\gamma _1}{{\bar \beta }^1} - {\beta _1}{{\bar \gamma }^1} - \tfrac{d}{2} + 1=\N_1-\tfrac{d}{2} + 1,\\
&{\J^2_2} \equiv -{\J_{24}} = {\theta _2}{{\bar \theta }^2} + {\gamma _2}{{\bar \beta }^2} - {\beta _2}{{\bar \gamma }^2} - \tfrac{d}{2} + 1=\N_2-\tfrac{d}{2} + 1,\\
&\mathcal{Y} \equiv -\frac{1}{2}\left( {i\left( {{\J_{12}} + {\J_{34}}} \right) + {\J_{14}} + {\J_{23}}} \right) = {\theta _1}{{\bar \theta }^2} + {\gamma _1}{{\bar \beta }^2} - {\beta _1}{{\bar \gamma }^2},\\
&\mathcal{Y}^\dagger \equiv -\frac{1}{2}\left( { - i\left( {{\J_{12}} + {\J_{34}}} \right) + {\J_{14}} + {\J_{23}}} \right) = {\theta _2}{{\bar \theta }^1} + {\gamma _2}{{\bar \beta }^1} - {\beta _2}{{\bar \gamma }^1},\\
&\gTr \equiv -\frac{1}{2}\left( {i{\J_{12}} - i{\J_{34}} - {\J_{23}} + {\J_{14}}} \right) = {{\bar \theta }^1}{{\bar \theta }^2} - {{\bar \beta }^1}{{\bar \gamma }^2} + {{\bar \beta }^2}{{\bar \gamma }^1},\\
&{\cal G} \equiv -\frac{1}{2}\left( {i{\J_{12}} - i{\J_{34}} + {\J_{23}} - {\J_{14}}} \right) = {\theta _1}{\theta _2} - {\beta _1}{\gamma _2} + {\beta _2}{\gamma _1}.
\end{split}
\end{equation}
The commutation relations among the new generators are,
\begin{equation}
\begin{split}
&[\gTr,{\cal G}]= \N_1+\N_2-d+2, \ \  [\N_1,\mathcal{Y}]=\Y, \ \ [\N_2,\mathcal{Y}^\dagger]=\Y^\dagger, \ \ [\mathcal{Y},\mathcal{Y}^\dagger] = \N_1 - \N_2, \\
&[\gTr,\N_1]=[\gTr,\N_2]=\gTr, \ \ [\N_1,{\cal G}]=[\N_2,{\cal G}]={\cal G},
\end{split}
\end{equation}
where all the other commutation relations vanish.

\section{From string to particle - NS sector}\label{string}
We consider the reduction from the Polyakov string to the point particle. Since the point particle admits the NS spectrum we consider the Polyakov action in the NS-NS sector. We start from the world sheet action with $z=e^{-i\omega}$, where $\omega = \sigma^1 + i\sigma^2$,
\begin{equation}
\begin{split}
    \sqrt{\alpha'}X^\mu\left(z,\bar{z}\right) &= x^\mu_0 - i\frac{\alpha'}{2}p^\mu \ln{|z|^2} + i\left(\frac{\alpha'}{2}\right)^{\frac{1}{2}}\sum_{m\in \mathcal{Z}'}\frac{1}{m}\left(\frac{\alpha^\mu_m}{z^m}+\frac{\tilde{\alpha}^\mu_m}{\bar{z}^m}\right) \\
    &\equiv \frac{\alpha'}{\sqrt{2}}\phi^\mu\left(\sigma^2\right) + i\left(\frac{\alpha'}{2}\right)^{\frac{1}{2}}\sum_{m\in \mathcal{Z}'}\frac{1}{m}\left(\alpha^\mu_m\left(\sigma^2\right)e^{im\sigma^1}+\tilde{\alpha}^\mu_m\left(\sigma^2\right)e^{-im\sigma^1}\right).
\end{split}
\end{equation}
The NS fermion has no zero mode on the cylinder due to its antisymmetry. To circumvent this problem we perform a twisted compactification. One way to do this is to start with the expansion of $\psi$ on the complex plane, that is,
\begin{equation}\label{fermionexpansion}
\begin{split}
\psi^\mu \left(z\right) = \sum_{r \in \mathcal{Z} + \frac{1}{2}}\frac{\psi^\mu_r}{z^{r+1/2}}, \hspace{20pt} \tilde{\psi}^\mu \left(\bar{z}\right) = \sum_{r \in \mathcal{Z} + \frac{1}{2}}\frac{\tilde{\psi}^\mu_r}{\bar{z}^{r+1/2}},
\end{split}
\end{equation}
The extra $\frac{1}{2}$ in the exponent in the expansion of the fermion in (\ref{fermionexpansion}) comes from the transformation from the cylinder to the complex plane, since $\psi$ takes values in the spinor bundle, $K^\frac{1}{2}$. To continue, twist $\psi$ (and $\tilde{\psi}$) so that $\psi$ becomes a scalar on the world sheet. After mapping back to the cylinder we then have,
\begin{equation}\label{fermion_expansion}
\begin{split}
\psi^\mu \left(\sigma^1,\sigma^2\right) = \sum_{r \in \mathcal{Z} + \frac{1}{2}}\psi^\mu_r\left(\sigma^2\right)e^{i(r+1/2)\sigma^1}, \hspace{20pt} \tilde{\psi}^\mu \left(\sigma^1,\sigma^2\right) = \sum_{r \in \mathcal{Z} + \frac{1}{2}}\tilde{\psi}^\mu_r\left(\sigma^2\right)e^{-i(r+1/2)\sigma^1}.
\end{split}
\end{equation}
This twist is also indicated by the fact that in the worldline formulation $\psi$ is viewed as a scalar. Note, that the twist breaks the world sheet diffeomorphisms which mix the $\sigma^1$ and $\sigma^2$ directions. We allow this diffeomorphism breaking since in the worldline the $\sigma^1$ direction is absent. The reduced action then becomes,
\begin{equation}
\begin{split}
S &= \frac{1}{4\pi}\int\left(\frac{2}{\alpha'}\partial X^\mu \bar{\partial}X_\mu + \psi^\mu\bar{\partial}\psi_\mu + \tilde{\psi}^\mu\partial\tilde{\psi}_\mu \right)dzd\bar{z} \\
&=\frac{i}{2}\int \left(\partial_2\phi^\mu\partial_2\phi_\mu + e^{\sigma^2}\psi^\mu_{1/2}\partial_2\psi_{\mu -1/2} + e^{\sigma^2}\tilde{\psi}^\mu_{1/2}\partial_2\tilde{\psi}_{\mu -1/2}\right)d\sigma^2 + \dots
\end{split}
\end{equation}
where the ellipsis indicates higher excitation modes and $\partial_2$ means a derivative with respect to $\sigma^2$. Note that while $\psi_{1/2}$ is the hermitian conjugate of $\psi_{-1/2}$ w.r.t the bpz inner product it is not the hermitian conjugate with respect to the natural inner product for the reduced action. We then define,
\begin{equation}
\begin{split}
&\theta^\mu_1 \equiv \psi^\mu_{-1/2}, \ \ \ \ \ \ \theta^\mu_2 \equiv \tilde{\psi}^\mu_{-1/2}, \\
&\bar{\theta}^\mu_1 \equiv  i e^{\sigma^2}\psi^\mu_{1/2}, \ \ \  \bar{\theta}^\mu_2 \equiv  i e^{\sigma^2}\tilde{\psi}^\mu_{1/2},
\end{split}
\end{equation}
and changing $\sigma^2=i\tau$ we find, for the lowest excitation modes,
\begin{equation}
\begin{split}
S =\frac{1}{2}\int \left(\partial_\tau\phi^\mu\partial_\tau\phi_\mu + \bar{\theta}^{\mu i}\partial_\tau\theta_{i\mu}\right)d\tau,
\end{split}
\end{equation}
where $i=1,2$.

\bibliographystyle{unsrt}
\bibliography{ref.bib}
\end{document}